\documentclass[letterpaper, 10 pt, conference]{ieeeconf}

\usepackage{graphicx}
\usepackage{epstopdf}
\usepackage{epsfig}
\usepackage{times}
\usepackage{amsmath}
\usepackage{amssymb}
\usepackage{latexsym}
\usepackage{color}
\usepackage{verbatim}
\usepackage{cite}
\usepackage{BernsteinStyle2}
\usepackage[font = footnotesize]{caption}
\usepackage[font = footnotesize]{subcaption}
\usepackage[hidelinks]{hyperref}
\usepackage{diagbox}
\usepackage{float}

\newcommand\norm[1]{\left\lVert#1\right\rVert}

\usepackage{cancel}

\DeclareSymbolFont{matha}{OML}{txmi}{m}{it}% txfonts
\DeclareMathSymbol{\varv}{\mathord}{matha}{118}

\usepackage{makecell}

\makeatletter
\newcommand*\bigcdot{\mathpalette\bigcdot@{.5}}
\newcommand*\bigcdot@[2]{\mathbin{\vcenter{\hbox{\scalebox{#2}{$\m@th#1\bullet$}}}}}
\makeatother

\IEEEoverridecommandlockouts

\title{\LARGE
An Adaptive Kalman Filter that Learns \\ the Coloring Dynamics of the Process Noise 
}

\author{Mohammad Almuhaihi, and Dennis S. Bernstein%
\thanks{$^{*}$Mohammad Almuhaihi and Dennis S. Bernstein are with the Department of Aerospace Engineering, University of Michigan, Ann Arbor, MI 48109, USA 
{\tt\small muhaihi@umich.edu}}%
}

\begin{document}

\maketitle
\thispagestyle{empty}
\pagestyle{empty}

%%%%%%%%%%%%%%%%%%%%%%%%%%%%%%%%%%%%%%%%%%%%%%%%%%%%%%%%%%%
\begin{abstract}
In many applications of state estimation, the process noise is colored;  this case is addressed by applying the standard Kalman filter (KF) to dynamics that are augmented with the coloring dynamics.
The present paper considers the case where the coloring dynamics are unknown, which renders the estimates obtained from the standard approach suboptimal. 
To address this problem, the present paper proposes an adaptive technique based on the principle that, if the measurement noise is white, then the innovations sequence is white if and only if the process noise is white. 
Leveraging this fact, an Innovations-Whitening Adaptive Kalman Filter (IWAKF) is developed, which learns the process-noise coloring online. 
By embedding an unknown coloring filter in a state-augmentation framework, IWAKF adapts its parameters by minimizing the empirical autocorrelation of the innovations, thereby driving them toward whiteness and restoring near-optimality without prior knowledge of the coloring dynamics.
\end{abstract}

%and producing correlated innovations. 

%%%%%%%%%%%%%%%%%%%%%%%%%%%%%%%%%%%%%%%%%%%%%%%%%%%%%%%%%%%
\section{INTRODUCTION}

In the standard Kalman filter, the process noise is assumed to be white
\cite{jazwinski1970stochastic,crassidis2004optimal,simon2006optimal,anderson2013optimal}.
In many applications of state estimation, however, the process noise is colored.
This case can addressed by applying the standard Kalman filter (KF) to dynamics that are augmented with the coloring dynamics
\cite{anderson2013optimal}.
This technique provides optimal state estimates for the given coloring dynamics.
In practice, however, the process-noise spectrum is often colored with unknown coloring dynamics.
The present paper focuses on this scenario.

To address the problem of state estimation with unknown coloring dynamics, the present paper proposes an adaptive technique based on the principle that, if the measurement noise is white, then the innovations sequence is white if and only if the process noise is white. 
Leveraging this fact, an Innovations-Whitening Adaptive Kalman Filter (IWAKF) is developed, which learns the process-noise coloring online. 
By embedding an unknown coloring filter in a state-augmentation framework, IWAKF adapts its parameters by minimizing the empirical autocorrelation of the innovations, thereby driving them toward whiteness and restoring near-optimality without prior knowledge of the coloring dynamics.

%% Beginning of Lit Review
Recent efforts handle colored disturbances by either learning a disturbance model from covariance data or by de-correlating measurements in time. 
Zare \emph{et al} formulate a data-enhanced approach that recovers a low-complexity colored-input model via convex covariance completion and then realizes an innovations model for use with an Ensemble Kalman Filter (EnKF) \cite{zare2021data}. 
While effective at matching second-order statistics, this method relies on accurate batch/empirical covariances and requires solving a nuclear-norm–regularized optimization, which adds computational overhead and reduces responsiveness to time-varying spectra.

The use of the innovations process for adaptivity traces back to the seminal study \cite{mehra1970identification}. Under LTI dynamics and steady-state operation, Mehra shows that innovations whiteness is a necessary and sufficient signature of optimality and uses the sample autocorrelation of the innovations to (i) test optimality and (ii) estimate $R$ together with (parts of) $Q$ consistently from data. However, this identification proceeds at the level of the covariances $(Q,R)$ rather than through an explicit dynamical shaping model for process noise. Consequently, it is not designed to recover or track unknown coloring dynamics in real time. The present work retains the innovations–whiteness principle but shifts the focus from covariance tuning to online learning of a parametric coloring filter embedded within a state-augmentation architecture, adjusting its parameters to drive empirical innovations autocorrelations toward zero and thereby restore near-optimal performance without prior spectral knowledge. This bridge from Mehra’s optimality test to real-time identification closes the gap for colored process noise and motivates the development that follows.

%% Enf of Lit Review

The contents of the paper are as follows.
Section II formulates the problem, and presents a 2nd-order uncertainty coloring filter and the resulting augmented dynamics.
Assuming that the process noise and measurement noise are mutually independent and the measurement noise is white, Section III relates the the whiteness of the innovations and the whiteness of the process noise.
Section IV presents a numerical example involving a damped pendulum with three distinct coloring filters.
Finally, Section V provides some future research directions.

%%%%%%%%%%%%%%%%%%%%%%%%%%%%%%%%%%%%%%%%%%%%%%%%%%%%%%%%%%%
%\clearpage

\section{Problem Statement} \label{proplem_statement}
We consider the discrete-time state–space model
\begin{equation}\label{eq: system}
\begin{aligned}
    x_{k} &= A x_{k-1} + B v_{k-1}, \\ 
    {y}_k &= C x_k + n_k,
\end{aligned}
\end{equation}
where the dynamics matrix $A\in \mathbb{R}^{n\times n}$, the input matrix $B \in \mathbb{R}^{n\times l}$, and the output matrix $C \in \mathbb{R}^{m\times n}$. 
The process noise $v_k \sim \mathcal{N}(0,Q)$ and measurement noise $n_k \sim \mathcal{N}(0,R)$ are zero mean and mutually independent.
The standard Kalman filter (KF) recursions are summarized below for reference.

% Forecast (fc)
\noindent \textit{Forecast (fc)}
\begin{align}
x_{k}^{\mathrm{fc}} &= A\,x_{k-1}^{\mathrm{da}} + B\,u_{k-1}, \label{eq:fc_state}\\
e_{k}^{\mathrm{fc}} :&= x_{k} - x_{k}^{\mathrm{fc}}, \label{eq:fc_err} \\
P_{k}^{\mathrm{fc}} &= A\,P_{k-1}^{\mathrm{da}}\,A^{\top} + Q. \label{eq:fc_cov}
\end{align}

% Innovations and its covariance
\noindent \textit{Innovations and covariance}
\begin{align}
z_{k} &:= y_{k} - C\,x_{k}^{\mathrm{fc}} = C e_{k}^{\mathrm{fc}} + n_k, \label{eq:innov}\\
S_{k} &:= \mathrm{Cov}(z_{k}) = C_{k} P_{k}^{\mathrm{fc}} C_{k}^{\top}
  + R. \label{eq:innov cov}
\end{align}

% Data assimilation (da)
\noindent \textit{Data assimilation (da)}
\begin{align}
K_{k} &= P_{k}^{\mathrm{fc}} C^{\top} S_{k}^{-1}, \label{eq:gain}\\
x_{k}^{\mathrm{da}} &= x_{k}^{\mathrm{fc}} + K_{k}\,z_{k}, \label{eq:da_state}\\
e_{k}^{\mathrm{da}} :&= x_{k} - x_{k}^{\mathrm{da}}, \label{eq:da_err} \\
P_{k}^{\mathrm{da}} &= \big( I_{n\times n} - K_{k} C\big) P_{k}^{\mathrm{fc}}.  \label{eq:da_cov}
\end{align}
These equations are optimal under the classical assumption that both $v_k$ and $n_k$ are white (that is, i.i.d.) sequences. In the optimal KF, the innovations $z_k$ is white; in particular, its first-lag autocorrelation implied by \eqref{eq:innov} is zero. When the process noise is \emph{colored}, this whiteness property no longer holds and the KF becomes suboptimal.

%@MA:  There might be a difference between residual and innovation, but I am not sure what it is

%@DB: The difference I believe is in the second term, the innovations subtract the measurment from the prediction (before data assimilation) while the residual subtracts the measurment from KF estimate (after data assimilation).

%OK, thanks

\subsection{Coloring Filter}
The standard approach for the case of colored process noise is to augment the system with a coloring filter that captures the autocorrelation of $v_k$. As a representative model, consider the second-order strictly proper transfer function
\begin{align}
    H(\bfq)    = \frac{\gamma_3 \bfq + \gamma_2}{\bfq^2+ \gamma_1 \bfq + \gamma_0}, 
    \label{eq: shaping_TF}
\end{align}
where $\bfq$ is the forward shift operator and $\gamma_{0,1,2,3}$ are constants. In practice, however, these coloring dynamics are rarely known a priori. The objective of this article is therefore to develop an online algorithm that adapts to the unknown process-noise coloring in real time, thereby restoring near-optimality by driving the innovations towards whiteness.

\subsection{State Augmentation}
Consider the coloring filter realization for the process noise given by
\begin{equation}\label{eq: shaping_real}
\begin{aligned}
    X_{H,k+1}   &= A_H X_{H,k} + B_H w_k, \\
    v_k         &= C_H X_{H,k}, 
\end{aligned}
\end{equation}
where $w_k$ is white with distribution $\mathcal{N}\!\left(0,\frac{\sigma_v^2}{|H(z)|^2}\right)$. The controllable-canonical realization of the second-order strictly proper transfer function in \eqref{eq: shaping_TF} is given by
\begin{align*}
    A_H &= \begin{bmatrix}
                -\gamma_1  & -\gamma_0 \\
                1          & 0
            \end{bmatrix}, \quad 
    B_H = \begin{bmatrix} 1 \\ 0 \end{bmatrix}, \\
    C_H &= \begin{bmatrix} \gamma_2 & \gamma_3 \end{bmatrix}.
\end{align*}

To embed the coloring filter \eqref{eq: shaping_real} into the plant model \eqref{eq: system}, we write
\begin{align*}
    x_{k} &= A x_{k-1} + B v_{k-1} \\
            &= A x_{k-1} + B (C_H X_{H,k-1}) \\
            &= A x_{k-1} + B C_H (A_H X_{H,k-2}+B_H w_{k-2}),
\end{align*}
leading to the augmented system
\begin{equation}\label{eq: shaping_real2}
\begin{aligned}
    x_{{\rm aug},k} &= A_{\rm aug} x_{{\rm aug},k-1} + B_{\rm aug}  w_{k-1}, \\
    y_{{\rm aug},k}   &= C_{\rm aug} x_{{\rm aug},k}, 
\end{aligned}
\end{equation}
where
\begin{table}[H]
    \centering
        \begin{tabular}{ c c }
         $x_{{\rm aug},k} = \begin{bmatrix}x_k \\ x_{H,k} \end{bmatrix}$  ,
         &
         $
             A_{\rm aug} = 
             \begin{bmatrix}
                 A & B  C_H \\
                 0 & A_H
             \end{bmatrix}
         $
         \\ \\
         $
             B_{\rm aug} = \begin{bmatrix}
                 0_{n\times1}  \\ B_H
             \end{bmatrix}
         $  ,
         
         &
         $
         C_{\rm aug} = [C \quad 0_{1\times m}]
         $.
        \end{tabular}
\end{table}
This technique fully specifies the augmentation of the coloring filter into the system dynamics. If the coloring model matches the true coloring of $v_k$, then the KF applied to the augmented system remains optimal and the innovations is white (that is, its first-lag autocorrelation is zero). In practice, the coloring dynamics may not be known exactly, which induces residual autocorrelation and suboptimality. Accordingly, we introduce an online optimization procedure to learn the coloring filter in real time.

%%%%%%%%%%%%%%%%%%%%%%%%%%%%%%%%%%%%%%%%%%%%%%%%%%%%%%%%%%%
\section{Whiteness of Process Noise and Innovations}
This section establishes an equivalence between the whiteness of the innovations and the whiteness of the process noise under the assumption that the process noise $v_k$ and measurement noise $n_k$ are mutually independent and measurement noise $n_k$ is white.

{\bf Proposition 1.}
Assume that the process noise $v_k$ and measurement noise $n_k$ for a linear time-invariant (LTI) system are mutually independent and measurement noise $n_k$ is white.
Then,
\begin{equation}
\Gamma_z(\ell)=0\ \ \forall\,\ell\ge1
\end{equation}
if and only if
\begin{equation}
    \Gamma_v(\ell)=0\ \ \forall\,\ell\ge1,
\end{equation}
where 
\begin{align*}
    \Gamma_z(\ell) \triangleq \mathrm{cov}(z_k,z_{k-\ell}), \quad \Gamma_v(\ell) \triangleq \mathrm{cov}(v_k,v_{k-\ell})
\end{align*}

%That is, the innovations sequence is white if and only if $v_k$ is white, provided $n_k$ is white. 

\begin{proof}
We rewrite the data assimilation step by substituting \eqref{eq:innov} into \eqref{eq:da_state}, which yields
\begin{align*}
    x_{k}^{\mathrm{da}} &= x_{k}^{\mathrm{fc}} + K_{k}\,(C e_{k}^{\mathrm{fc}} + n_k) .
\end{align*}
Substituting back into \eqref{eq:da_err} yields
\begin{align*}
    e_{k}^{\mathrm{da}} &= x_{k} - x_{k}^{\mathrm{da}} \nonumber \\
    &= x_{k} - x_{k}^{\mathrm{fc}} - K_{k}\,(C e_{k}^{\mathrm{fc}} + n_k) \nonumber\\
    &= e_{k}^{\mathrm{fc}} - K_{k}\,(C e_{k}^{\mathrm{fc}} + n_k) \nonumber\\
    &= (I-K_k C)e_{k}^{\mathrm{fc}} - K_k n_k.
\end{align*}
Propagating to the next forecast error yields
\begin{align*}
    e_{k+1}^{\mathrm{fc}} 
    &= A e_{k}^{\mathrm{da}} + v_{k+1} \nonumber\\
    &= A\!\left( (I-K_k C)e_{k}^{\mathrm{fc}} - K_k n_k \right) + v_{k+1} \nonumber \\
    &= A(I-K_kC)\,e_k^{\mathrm{fc}} + v_{k+1} - A K_k n_k \nonumber \\
    &= F_k\,e_k^{\mathrm{fc}} + v_{k+1} - A K_k n_k,
\end{align*}
where $F_k = A(I-K_kC)$, which is assumed to be stable, that is, $\rho(F)<1$. 
%
%@MA:  Need to be careful here since you have a time varying system and stability is not assured by the spectral radius
%my advice is to confine the proposition to the case where A and C are constant---that is, LTV dynamics
%
%@DB: You mean to limit our proposition to cover only LTI systems? 
%Yes, for the proposition, but you can still consider LTV systems
%
% @DB: Done
Then $e_{k}^{\mathrm{fc}}$ is unfolded under steady state as
\begin{align*}
    e_{k}^{\mathrm{fc}} &= \sum_{i=1}^{\infty} F^{i-1} (v_{k-i+1}-A K n_{k-i})
\end{align*}
Substituting back into \eqref{eq:innov} gives
\begin{align*}
    z_k = n_k + \sum_{j=0}^{\infty} C F^{j} v_{k-j} - \sum_{i=1}^{\infty} C F^{i-1} A K n_{k-i}
\end{align*}
Define
\begin{align}
    \mathcal{G}(\bfq^{-1}) &\triangleq \sum_{i=1}^{\infty} C F^{i-1} A K \bfq^{-i} \\
    \mathcal{H}(\bfq^{-1}) &\triangleq \sum_{j=0}^{\infty} C F^{j} \bfq^{-j}
\end{align}
where $\bfq^{-1}$ is the backward-shift operator. Rewriting yields
\begin{align*}
    z_k = (I-\mathcal{G}(\bfq^{-1}))n_k + \mathcal{H}(\bfq^{-1}) v_k
\end{align*}
Transforming to the frequency domain gives
\begin{align*}
    Z(e^{\jmath\phi}) = (I-G(e^{\jmath\phi}))N(e^{\jmath\phi}) + H(e^{\jmath\phi}) v_k,
\end{align*}
where
\begin{align}
    G(e^{\jmath\phi}) &= C(e^{\jmath\phi}I-F)^{-1} AKe^{-jw}, \\
    H(e^{\jmath\phi}) &= C(e^{\jmath\phi}I-F)^{-1}.
\end{align}
Using linear spectral mapping, the Power Spectral Density (PSD) of the signal $z$ can be computed as
\begin{align*}
    \Phi_z(\phi) = T_n \phi_\rmn(\phi) T_n^* + T_v \Phi_v(\phi) T_v^* 
\end{align*}
and the cross-correlation between $n$ and $v$ is zero since independence between process and measurement noise is assumed.
Since $n_k$ is white, it has a PSD of $R$ and the terms $T_n$ and $T_v$ are $(I-G)$ and $H$, respectively. Thus,
\begin{align}
    \Phi_z(\phi) = (I-G) R (I-G)^* + H \Phi_v(\phi) H^*.
    \label{eq:PSD of z}
\end{align}
Assuming that $\Phi_v(\phi)$ is white with covariance $Q$ yields
\begin{align*}
    \Phi_z(\phi) = (I-G) R (I-G)^* + H Q H^*,
\end{align*}
which implies that $\Phi_z(\phi)$ is constant across all frequencies, thereby ensuring that the signal $z$ is white. This establishes sufficiency.

To prove necessity, let $\Phi_z(\phi)=S$. Substituting back into \eqref{eq:PSD of z} yields
\begin{align*}
    S = (I-G) R (I-G)^* + H \Phi_v(\phi) H^*.
\end{align*}
Since $(C,F)$ is observable, which is implied by $(A,C)$ detectability, $H$ has full column rank. 
%
%@MA:  There is a problem here due to time varying dynamics
%
%
%
%
Thus, its pseudo-inverse $H^\dag$ exists, resulting in
\begin{align*}
    \Phi_v(\phi) = H^\dag (S - (I-G) R (I-G)^*) H^{\dag*}
\end{align*}
which holds only if $\Phi_v(\phi)$ is constant across all frequencies, which is the definition of whiteness. 
\end{proof}
%%%%%%%%%%%%%%%%%%%%%%%%%%%%%%%%%%%%%%%%%%%%%%%%%%%%%%%%%%%
\subsection{Innovations-Whitening Adaptive Kalman Filter (IWAKF)}
After proving whiteness duality between process noise and innovation, we leverage this relation by estimating the coloring filter characteristics through minimizing the autocorrelation in the innovation. In other words,
perfect state augmentation results in a white innovations sequence. The Innovations-Whitening Adaptive Kalman Filter (IWAKF) utilizes this fact to adapt to the coloring dynamics of process noise. Let 
\begin{align}
    J \triangleq \norm{\mathbb{E}[ z_{{\rm Aug.},k} z_{{\rm Aug.},k-\ell}]}
\end{align}
where $z_{{\rm Aug.},k}$ is the innovations of the augmented system and its autocorrelation is used as the cost function $J$. The goal is to find $\gamma_{0,1,2,3}$ that minimizes $J$
\begin{align}
    \gamma_{0,1,2,3} = \arg \min_{\gamma_{0,1,2,3}} J.
\end{align}

\section{Simulation and Performance}

%@MA:  I used roman c for continuous time dynamics.  Please explain to the reader what the dynamics represent

%phi_n and phi_d are not defined
%sine and cosine might confuse some readers so explain the notation
%define T

%@DB: I have redescribed the pendulim equations in continuous-time instead. phi_n and zeta are now defined. no need to define phi_d now as it is no longer shown explicitly. 

\begin{figure}
    \centering
    \includegraphics[width=0.5\linewidth]{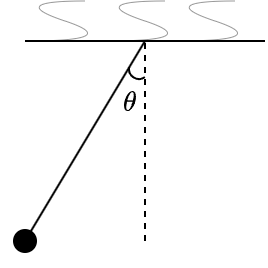}
    \caption{Pendulum schematic defining the angle $\theta$ and its reference.}
    \label{fig:pendulimFigure}
\end{figure}
\begin{figure}
    \centering
    \includegraphics[width=1\linewidth]{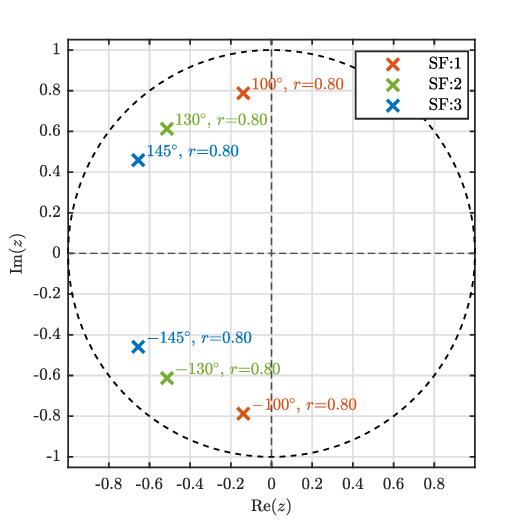}
    \caption{Coloring filter pole locations. The three coloring filters (SF:1, SF:2, SF:3) are characterized by their poles in the complex z-plane. All filters have poles at radius 0.8 with varying angles (100, 130, and 145 degrees), representing different coloring characteristics of the process noise.}
    \label{fig:SFs in complex plane}
\end{figure}

In the simulation, a damped-pendulum, shown in figure\ref{fig:pendulimFigure}, model is employed, using a state representation where the states are $\theta$ and $\dot \theta$, in accordance with \eqref{eq: system}. The equations are expressed as
\begin{equation}
\begin{aligned}
% A &= e^{-\zeta \phi_\rmn T}A_\rmc, \\
A &= e^{A_\rmc T}, \\
B &= \begin{bmatrix}
    0 \\ 1
\end{bmatrix} , \quad
C = \begin{bmatrix}
    1 & 0
\end{bmatrix}
\end{aligned}
\end{equation}
where
\begin{align*}
% A_\rmc = 
% \begin{bmatrix}
% \textbf{c}(\phi_\rmd T) + \frac{\zeta \phi_\rmn}{\phi_\rmd}\textbf{s}(\phi_\rmd T) & \frac{1}{\phi_\rmd}\textbf{s}(\phi_\rmd T) \\[10pt]
% -\frac{\phi_\rmn^2}{\phi_\rmd}\textbf{s}(\phi_\rmd T) & \textbf{c}(\phi_\rmd T) - \frac{\zeta \phi_\rmn}{\phi_\rmd}\textbf{s}(\phi_\rmd T)
% \end{bmatrix}\\
A_\rmc &= 
\begin{bmatrix}
0 & 1 \\[10pt]
- \phi_\rmn & -2\zeta \phi_\rmn
\end{bmatrix}\\
% \phi_\rmn = \sqrt{\frac{g}{L}}, \quad 
% \zeta &= \frac{b}{2 m L^{2} \phi_\rmn} \quad  
% \phi_\rmd = \phi_\rmn \sqrt{1 - \zeta^2}.
\phi_\rmn &= \sqrt{\frac{g}{L}}, \quad 
\zeta = \frac{b}{2 m L^{2} \phi_\rmn} .
\end{align*}
$\phi_n$ and $\zeta$ are the undamped natural frequency and damping ratio, respectively.
The colored process noise is shaped by \eqref{eq: shaping_TF} with three distinct coloring filters defined in Table \ref{table: list of SFs} and Figure \ref{fig:SFs in complex plane}. 
\begin{table}[]
    \centering
    \begin{tabular}{c | c | c}
        Shortcut & Poles & Zeros \\ \hline
        SF:1 & $0.8e^{j\pm100}$ & $3$ \\
        SF:2 & $0.8e^{j\pm130}$ & $3$ \\
        SF:3 & $0.8e^{j\pm145}$ & $3$ \\
    \end{tabular}
    \caption{Coloring filter specifications. The table lists the pole and zero locations for three example coloring filters. Each filter has conjugate poles at radius 0.8 with angles of 100, 130, and 145 degrees, and a single zero at 3.}\label{table: list of SFs}
\end{table}

\begin{figure}
    \centering
    \includegraphics[width=1\linewidth]{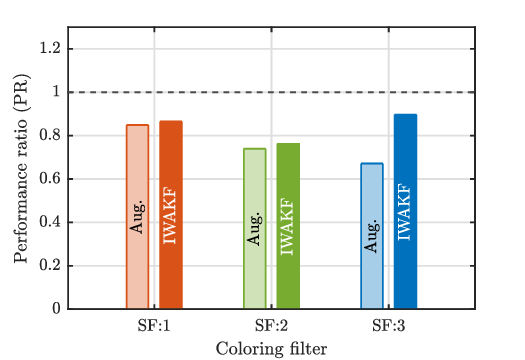}
    \caption{Performance ratio comparison. The figure compares the performance ratios $\rm{PR}_{Aug.}$ (perfectly augmented KF with known coloring dynamics) and $\rm{PR}_{IWAKF}$ (Innovations-Whitening Adaptive KF) for three coloring filter examples (SF:1, SF:2, SF:3). Both methods achieve similar performance ratios, demonstrating that IWAKF can match the performance of the perfectly augmented filter without prior knowledge of the process-noise coloring dynamics.}
    \label{fig:performance PR}
\end{figure}

\begin{figure}
    \centering
    \includegraphics[width=1\linewidth]{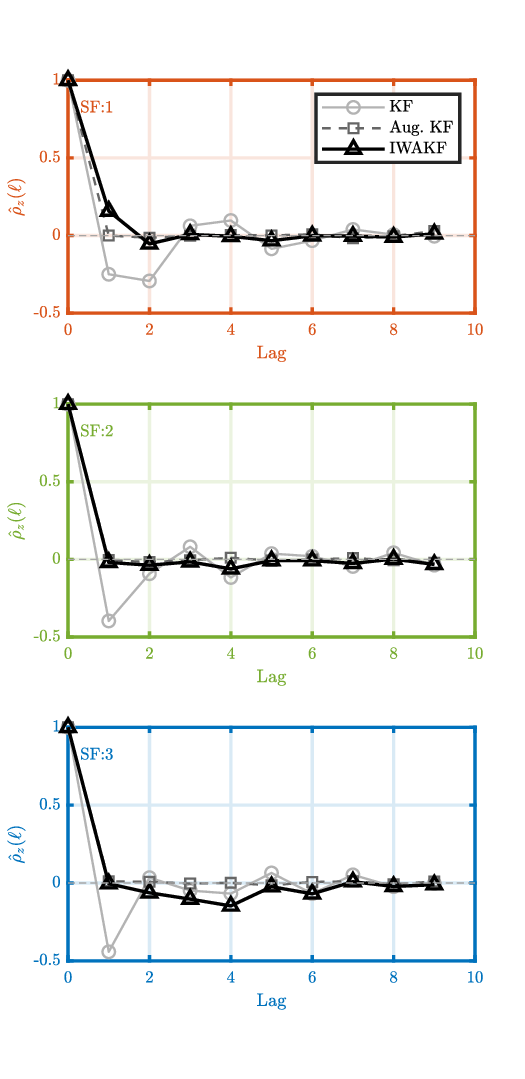}
    \caption{Innovations autocorrelation comparison. The figure displays the autocorrelation of the innovations sequence for the standard KF, the augmented KF (Aug. KF), and IWAKF. The standard KF shows significant autocorrelation, indicating colored innovations due to unmodeled colored process noise. Both the augmented KF and IWAKF exhibit near-zero autocorrelation, demonstrating that they successfully whiten the innovation. Notably, IWAKF achieves this whitening without prior knowledge of the process-noise coloring dynamics.}
    \label{fig:performance AC}
\end{figure}

For each transfer function, a performance comparison is made between the IWAKF and the perfectly augmented Kalman filter where the coloring dynamics of the process noise coloring are known for the KF. Moreover, for each case, the Performance Ratio (PR) is used for better comparison and is defined as
\begin{align}
    \rm{PR}_{Aug.} \triangleq \frac{\rm RMSE_{Aug.}}{\rm RMSE_{KF}}, 
\end{align}
and
\begin{align}
    \rm{PR}_{IWAKF} \triangleq \frac{\rm RMSE_{IWAKF}}{\rm RMSE_{KF}},
\end{align}
where 
\begin{align*}
    {\rm RMSE} = \sqrt{\frac{1}{N} \sum_{k=1}^{N} {e_k^{\rm da}}^{2}},
\end{align*}
and $\rm RMSE_{KF}$ is the error in the standard Kalman filter with no augmentation. Figure \ref{fig:performance PR} shows the performance of IWAKF compared with the perfectly augmented KF.
Figure \ref{fig:performance AC} demonstrates the innovations of KF, Aug.KF and IWAKF where both the Aug.KF and IWAKF successfully whitened the innovation, whereas the latter was accomplished with no previous knowledge of the exact coloring filter.
%%%%%%%%%%%%%%%%%%%%%%%%%%%%%%%%%%%%%%%%%%%%%%%%%%%%%%%%%%%
% \clearpage
\section{DISCUSSION AND FUTURE RESEARCH} \label{dis_con_fut}
This paper developed an Innovations-Whitening Adaptive Kalman Filter (IWAKF) that learns the coloring dynamics of the process noise online by minimizing the empirical autocorrelation of the innovations under white measurement noise. 
Simulations on a damped-pendulum example indicate that IWAKF attains performance comparable to a perfectly augmented KF that has complete knowledge of the true coloring filter, thereby restoring near-optimal estimation without prior spectral information.
The analysis herein relies on LTI assumptions to establish the equivalence between innovations whiteness and process-noise whiteness. A primary direction is to extend the theory and implementation to linear time-varying (LTV) systems with time-indexed dynamics and outputs, where conditions such as uniform detectability/stabilizability, bounded time-varying gains, and persistence of excitation replace their LTI counterparts. This requires a whiteness characterization that is valid under nonstationary spectra, together with finite-window identification tools and change detection to track drifting coloring dynamics.

\section*{ACKNOWLEDGMENTS}
The first author is grateful for a fellowship provided by Saudi Arabian Cultural Mission (SACM).

\nocite{chang2014kalman}
%\bibliography{bibpaper,bibijc}
\bibliography{bibAll}
\bibliographystyle{ieeetr}

\end{document}